\documentclass[showpacs,preprintnumbers,amsmath,amssymb,prb,10pt,aps]{revtex4-1}

\usepackage{graphicx}
\usepackage{dcolumn}
\usepackage{bm}

\begin{document}

\title{Anomalously large damping of long-wavelength quasiparticles caused by long-range interaction}

\author{A. V. Syromyatnikov}
 \email{syromyat@thd.pnpi.spb.ru}
\affiliation{Petersburg Nuclear Physics Institute, Gatchina, St.\ Petersburg 188300, Russia}
\affiliation{Department of Physics, St.\ Petersburg State University, 198504 St.\ Petersburg, Russia}

\date{\today}

\begin{abstract}

We demonstrate that long-range interaction in a system {\it can} lead to a very strong interaction between long-wavelength quasiparticles and make them heavily damped. In particular, we discuss magnon spectrum using $1/S$ expansion in 3D Heisenberg ferromagnet (FM) with arbitrary small dipolar forces at $T\ll T_C$. We obtain that a fraction of long-wavelength magnons with energies $\epsilon_{\bf k}<T$ has anomalously large damping $\Gamma_{\bf k}$ ($\Gamma_{\bf k}/\epsilon_{\bf k}$ reaches 0.3 for certain $\bf k$). This effect is observed both in quantum and classical FMs. Remarkably, this result contradicts expectation of the quasiparticle concept according which a weakly excited state of a many-body system can be represented as a collection of weakly interacting elementary excitations. Particular materials are pointed out which are suitable for corresponding experiments.

\end{abstract}


\maketitle

\section{Introduction}
\label{int}

The concept of elementary excitations or quasiparticles is one of the most powerful tools in discussion of low-energy properties of strongly interacting many-body systems. \cite{agd,stphys} According to this concept a weakly excited state of a system can be represented as a collection of propagating weakly interacting quasiparticles which carry quanta of energy $\epsilon_{\bf k}$ and momentum $\bf k$. Elementary excitations, being wave-packets of stationary states, have finite lifetime (or damping $\Gamma_{\bf k}$) which is interpreted as a result of quasiparticles spontaneous decay and interaction between them (at $T\ne0$). According to the quasiparticle concept supported by quite a general line of argument \cite{agd,stphys,hyd} (not considering, however, long-range interactions in a system), long-wavelength elementary excitations are well-defined (i.e., $\epsilon_{\bf k}\gg\Gamma_{\bf k}$). It is usually the case also for short-wavelength quasiparticles but a limiting number of systems is known in which their lifetime is very short or zero. Thus, the spectrum of short-wavelength elementary excitations in liquid $^4$He crosses a two-particle continuum at threshold momentum $k_c$. As a result spontaneous decay of an elementary excitation into two quasiparticles is allowed at $k>k_c$ by energy and momentum conservation laws written as \cite{pit,*he1,*he2}
\begin{equation}
\label{split}
\epsilon_{\bf k} = \epsilon_{\bf q} + \epsilon_{\bf k-q}.
\end{equation}
Decay processes are so strong in liquid $^4$He that $k=k_c$ is a termination point of the spectrum. A similar behavior of short-wavelength quasiparticles (magnons) has been observed recently in a number of magnetic systems: quasi-2D spin liquid \cite{liqinst,liqth1,*liqth2}, quasi-1D gapped spin system \cite{1dinst,liqth1,*liqth2} and in quasi-2D antiferromagnet (AF) with $S=5/2$ in strong magnetic field \cite{afexp}. The one-magnon branch disappears completely at $k>k_c$ in the quasi-1D material while the ratio $\Gamma_{\bf k}/\epsilon_{\bf k}$ amounts to 0.1 above the threshold in the quasi-2D systems. 

We demonstrate in the present paper that in contrast to the quasiparticle concept expectation even small long-range interaction in a system {\it can} lead to a very strong interaction between long-wavelength quasiparticles and make them badly defined. In particular, we show that a fraction of long-wavelength magnons with $\epsilon_{\bf k}<T$ is heavily damped ($\Gamma_{\bf k}/\epsilon_{\bf k}$ reaches 0.3 for certain $\bf k$) in 3D Heisenberg ferromagnet (FM) on a cubic lattice with arbitrary small dipolar forces at small temperature $T\ll T_C$ which Hamiltonian has the form 
\begin{equation}
\label{ham}
{\cal H} = -\frac12 \sum_{l\ne m} \left(J_{lm}\delta_{\rho\beta} + Q_{lm}^{\rho\beta}\right) S_l^\rho  S_m^\beta,
\end{equation}
where $Q_{lm}^{\rho\beta} = (g\mu)^2(3R_{lm}^\rho R_{lm}^\beta - \delta_{\rho\beta}R_{lm}^2)/R_{lm}^5$ is the dipolar tensor. Although this model was extensively studied before and it describes well a class of magnetic materials (see below), renormalization of the magnon spectrum by thermal and quantum fluctuations has not been analyzed thoroughly yet. It is well known that the spectrum in the linear spin-wave approximation (classical spectrum) at $L^{-1}\ll k\ll1$ has the form \cite{hp,sw}
\begin{eqnarray}
\label{spec0}
\epsilon_{0\bf k} &=& \sqrt{(Dk^2+g\mu H^{(i)})\left(Dk^2 + g\mu H^{(i)} + S\omega_0\sin^2\theta_{\bf k}\right)},\\
\label{o0}
\omega_0 &=& 4\pi (g\mu)^2,
\end{eqnarray}
where we set the lattice spacing equal to unity, $\omega_0$ is the characteristic dipolar energy, $D$ is the spin-wave stiffness which is equal to $SJ$ for cubic FM with exchange coupling $J$ between nearest neighbors only, $\theta_{\bf k}$ is the angle between magnetization and $\bf k$, $H^{(i)}$ is the intrinsic magnetic field which is zero, e.g., in the multidomain sample \cite{landau} and $L$ is the characteristic length of a given domain. We assume below that $T\ll SD\sim T_C$ (i.e., we do not consider frustration which can reduce $D$ considerably and assume that $T_C\sim SD\sim S^2J$) and $D\gg S\omega_0$ as it usually is. 

Long-wavelength magnons are well defined in the model \eqref{ham} with $Q_{lm}^{\rho\beta}=0$ at $T\ll T_C$ which damping has the form $\Gamma_{\bf k}\propto T^2k^4\ln k$. \cite{kash,*vaks,*harris,hyd} Notice that this previous finding is in full agreement with the quasiparticle concept because 3D FM is weakly excited at $T\ll T_C$. We demonstrate below that dipolar forces, despite their smallness and due to their long-range nature, give rise to great renormalization of the bare spectrum \eqref{spec0}. Renormalization of the real part of the spectrum has been discussed in our previous paper \cite{idipfm,*ierr}. We show there that fluctuations lead to the gap in the spectrum which resolve problems of infrared singularities of some observables obtained by other authors (see discussion below). We turn in the present paper to calculation of the magnon damping.

Dipolar forces lead to three-magnon vertexes giving rise to processes of magnon decay \eqref{split}. However, we obtain below that confluence processes reduce the magnon lifetime much stronger. They arise only at $T\ne0$ and have the following conservation law (cf.\ Eq.~\eqref{split}):
\begin{equation}
\label{conf}
\epsilon_{\bf k} = \epsilon_{\bf q} - \epsilon_{\bf k-q}.
\end{equation}
The important role of these processes for magnon relaxation was recognized long time ago in Refs.~\cite{shlom,*sparks}. These works were motivated by non-linear ferromagnetic resonance experiments. Therefore large external magnetic field was taken into account there leading to $g\mu H^{(i)}\agt S\omega_0$. It is seen from expressions obtained in Refs.~\cite{shlom} that $\epsilon_{\bf k}\gg\Gamma_{\bf k}$ at $g\mu H^{(i)}\sim S\omega_0$ but $\Gamma_{\bf k}\to\infty$ as $k,H^{(i)}\to0$ and results should be reconsidered if $H^{(i)}=0$. We perform this reconsideration below. 

We derive magnon damping using $1/S$ expansion. It is shown that in accordance with previous results \cite{shlom,*sparks} $\Gamma_{\bf k}$ diverges as $k\to0$ in the first order in $1/S$ that is a consequence of the Goldstone character of the bare spectrum \eqref{spec0} at $H^{(i)}=0$. This divergence signifies that the spectrum cannot be found using the conventional $1/S$ expansion. One notice, however, that this singularity should be screened by the gap in the spectrum. That is why the easiest way to find the spectrum in this case is to perform the self-consistent calculations. As a result we show that the main corrections to the spectrum comes from diagrams of the first order in $1/S$ whereas those from higher order ones are small by the parameter $S\omega_0/D\ll1$. We demonstrate below that magnons are well defined at $T=0$, while thermal fluctuations lead to a great enhancement of the damping: a peak arises in the ratio $\Gamma_{\bf k}/\epsilon_{\bf k}$ at small $k$ which height reaches 0.3 for momenta directed nearly along magnetization. The fraction of overdamped magnons is small and one could expect small influence from them on the system properties (magnetization, specific heat, etc.). By reducing dipolar forces radius of action we demonstrate that it is their long-range nature that is responsible for such a remarkable suppression of the long-wavelength quasiparticles. We show that this suppression can be seen both in quantum and classical FMs because thermal fluctuations are responsible for it. 

Interestingly, quantum and thermal fluctuations lead to smaller $\Gamma_{\bf k}/\epsilon_{\bf k}$ in lower dimension 2D FM on square lattice with dipolar forces that is discussed in our previous paper Ref.~\cite{i2d}. We find there that thermal fluctuations also lead to a peak in $\Gamma_{\bf k}/\epsilon_{\bf k}$ at small $k$ in 2D FM which height, however, is of the order of $T/T_C\ll1$ for $S\sim1$ and reaches the value of 0.16 for $S\gg1$. We show below that the origin of the greater role which play fluctuations in higher dimension is that dipolar forces give rise to larger three-particle vertexes in 3D FM.

The rest of the present paper is organized as follows. The Hamiltonian transformation and the technique are discussed in Sec.~\ref{hamtrans}. Renormalization of the energy and the real part of the spectrum are discussed briefly in Sec.~\ref{realp}. Magnon damping is considered in Sec.~\ref{md}. In Sec.~\ref{disc} we i) show that the observed anomaly in the damping can be seen both in quantum and classical FMs, ii) compare the damping in 2D and 3D FMs and make a counter intuitive conclusion that it is smaller (compared to the real part of the spectrum) in lower dimensional FM, and iii) discuss particular compounds suitable for corresponding experimental verification of our results in 3D FM. Sec.~\ref{con} contains our conclusion. An appendix is included with some details of calculations.

\section{Hamiltonian transformation}
\label{hamtrans}

After the Fourier transformation Hamiltonian \eqref{ham} is written as 
\begin{equation}
\label{hamk}
	{\cal H} = -\frac12 \sum_{\bf k}\left(J_{\bf k}\delta_{\alpha\beta} + Q_{\bf k}^{\alpha\beta}\right) S_{\bf k}^\alpha  S_{-\bf k}^\beta - g\mu H  \mathfrak N S_{\bf 0}^z,
\end{equation}
where $J_{\bf k} = \sum_l J_{lm}\exp(i{\bf k R}_{lm})$ and $Q_{\bf k}^{\alpha\beta} = \sum_l Q_{lm}^{\alpha\beta}\exp(i{\bf k R}_{lm})$. Notice that we take into account the Zeeman term $-g\mu H\sum_iS_i^z$ in Eq.~\eqref{hamk} that is necessary to do in order to describe the real finite size samples: $H$ is the field in a given domain produced by all other domains in the multidomain sample in zero external magnetic field. We will assume that $H=0$ in the unidomain sample that is magnetized in the direction in which it can be considered to be infinite. The dipolar tensor $Q_{\bf k}^{\alpha\beta}$ possesses the well-known properties \cite{hp,sw,sums}
\begin{eqnarray}
\label{qsmall}
Q_{\bf k}^{\alpha\beta} &=& \omega_0\left( \frac {\delta_{\alpha\beta}}{3} - \frac{k_\alpha k_\beta}{k^2} \right),  
\mbox{ if } \frac 1L\ll k\ll1,\\
Q_{\bf 0}^{\alpha\beta} &=& \omega_0 \left(\frac13 - {\cal N}_\alpha\right)\delta_{\alpha\beta},
\end{eqnarray}
where ${\cal N}_\alpha$ are demagnetizing factors. After Dyson-Maleev transformation
\begin{eqnarray}
S^x_{\bf k} &=& \sqrt{\frac S2} \left( a_{\bf k} + a^\dagger_{-\bf k} - \frac{(a^\dagger a^2)_{\bf k}}{2S} \right), \nonumber\\
S^y_{\bf k} &=& -i\sqrt{\frac S2} \left( a_{\bf k} - a^\dagger_{-\bf k} - \frac{(a^\dagger a^2)_{\bf k}}{2S}\right),\\
S^z_{\bf k} &=& S - (a^\dagger a)_{\bf k}
\nonumber
\end{eqnarray}
Hamiltonian \eqref{hamk} has the form ${\cal H} = E_0 + \sum_{i=1}^6 {\cal H}_i$, where $E_0$ is the classical ground state energy and ${\cal H}_i$ denote terms containing products of $i$ operators $a$ and $a^\dagger$. In particular, ${\cal H}_1=0$ because it contains only $Q_{\bf 0}^{\alpha\beta}$ with $\alpha\ne\beta$ and
\begin{eqnarray}
\label{h2}
{\cal H}_2 &=& \sum_{\bf k} \left[E_{\bf k} a^\dagger_{\bf k}a_{\bf k} + \frac{B_{\bf k}}{2} a_{\bf k}a_{-\bf k} + 
\frac{B_{\bf k}^*}{2} a^\dagger_{\bf k}a^\dagger_{-\bf k}\right],\\
{\cal H}_3 &=& \sqrt{\frac {S}{2 \mathfrak N}} \sum_{{\bf k}_1 + {\bf k}_2 + {\bf k}_3 = 0} a^\dagger_{-1}\left[a^\dagger_{-2}(Q_2^{xz}+iQ_2^{yz}) 
\right.\nonumber\\
\label{h3}
&&{}\left.
+ a_2(Q_2^{xz} - iQ_2^{yz})\right]a_3,\\
{\cal H}_4 &=& \frac{1}{4 \mathfrak N}\sum_{{\bf k}_1 + {\bf k}_2 + {\bf k}_3 + {\bf k}_4 = 0} 
\left\{
2(J_1-J_{1+3})a^\dagger_{-1}a^\dagger_{-2}a_3a_4 \right.\nonumber\\
&&{}
+ a^\dagger_{-1} \left[a_2 (Q_2^{xx} - 2iQ_2^{xy} - Q_2^{yy}) \right.\nonumber\\
\label{h4}
&&{}\left.\left.
+ a^\dagger_{-2}( Q_2^{xx} + Q_2^{yy} - 2Q_{2+3}^{zz}) \right] a_3a_4 \right\},
\end{eqnarray}
where $\mathfrak N$ is the number of spins in the lattice, we drop index $\bf k$ in Eqs.~(\ref{h3}) and (\ref{h4}) and \begin{eqnarray}
\label{e}
E_{\bf k} &=& S(J_{\bf 0} - J_{\bf k}) - \frac S2\left( Q_{\bf k}^{xx} + Q_{\bf k}^{yy} - \frac{2\omega_0}{3} \right) + g\mu(H - 4\pi g\mu S{\cal N}_z) \nonumber\\
& \stackrel{k\ll1}{\approx} & Dk^2 + \frac{S\omega_0}{2}\sin^2\theta_{\bf k} + g\mu(H - 4\pi g\mu S{\cal N}_z),
\\
\label{b}
B_{\bf k} &=& -\frac S2 \left( Q_{\bf k}^{xx} - 2iQ_{\bf k}^{xy} - Q_{\bf k}^{yy} \right)
\stackrel{k\ll1}{\approx} \frac{S\omega_0}{2}\sin^2\theta_{\bf k}e^{-2i\phi_{\bf k}},
\end{eqnarray}
where expressions after $\stackrel{k\ll1}{\approx}$ are approximate values of corresponding quantities at $L^{-1}\ll k\ll1$ and $\phi_{\bf k}$ is the azimuthal angle of $\bf k$. The expression in the brackets of the last term in Eq.~(\ref{e}) is the intrinsic magnetic field $H^{(i)}$. In the multidomain sample the term $4\pi g\mu S{\cal N}_z$ is the demagnetizing field of the considered domain that is equal to $H$ in the domain volume so that $H^{(i)}=0$. \cite{landau} The intrinsic field is zero also in a unidomain sample that is infinite in the direction of magnetization if the external field is zero because ${\cal N}_z=0$ in this case. One leads to Eq.~\eqref{spec0} from Eqs.~\eqref{h2}, \eqref{e} and \eqref{b} for the spectrum given in the linear spin-wave approximation by
\begin{equation}
\label{spec1}
\epsilon_{0\bf k} = \sqrt{E_{\bf k}^2 - |B_{\bf k}|^2}.
\end{equation}
 
To perform the calculations it is convenient to introduce the following retarded Green's functions: $G(\omega,{\bf k}) = \langle a_{\bf k}, a^\dagger_{\bf k} \rangle_\omega$, $F(\omega,{\bf k}) = \langle a_{\bf k}, a_{-\bf k} \rangle_\omega$, ${\overline G}(\omega,{\bf k}) = \langle a^\dagger_{-\bf k}, a_{-\bf k} \rangle_\omega = G^*(-\omega,-{\bf k})$ and $F^\dagger (\omega,{\bf k}) = \langle a^\dagger_{-\bf k}, a^\dagger_{\bf k} \rangle_\omega = F^*(-\omega,-{\bf k})$. We have two sets of Dyson equations for them. One of these sets has the form:
\begin{equation}
\label{eqfunc}
\begin{array}{l}
G(\omega,{\bf k}) = G^{(0)}(\omega,{\bf k}) + G^{(0)}(\omega,{\bf k}){\overline \Sigma}(\omega,{\bf k})G(\omega,{\bf k}) + G^{(0)}(\omega,{\bf k}) [B_{\bf k}^* + \Pi(\omega,{\bf k})] F^\dagger(\omega,{\bf k}),\\
F^\dagger(\omega,{\bf k}) = {\overline G}^{(0)}(\omega,{\bf k}) \Sigma(\omega,{\bf k})F^\dagger(\omega,{\bf k}) + {\overline G}^{(0)}(\omega,{\bf k}) [B_{\bf k} + \Pi^\dagger(\omega,{\bf k}) ]G(\omega,{\bf k}),
\end{array}
\end{equation}
where $G^{(0)}(\omega,{\bf k}) = (\omega - E_{\bf k}+i\delta)^{-1}$ is the bare Green's function and $\Sigma$, $\overline \Sigma$, $\Pi$ and $\Pi^\dagger$ are the self-energy parts. Solving Eqs.~(\ref{eqfunc}) one obtains:
\begin{eqnarray}
G(\omega,{\bf k}) &=& \frac{\omega + E_{\bf k} + \Sigma(\omega,{\bf k})}{{\cal D}(\omega,{\bf k})},\nonumber\\
F(\omega,{\bf k}) &=& -\frac{B^*_{\bf k} + \Pi(\omega,{\bf k})}{{\cal D}(\omega,{\bf k})},\nonumber\\
\label{gf}
{\overline G}(\omega,{\bf k}) &=& \frac{-\omega + E_{\bf k} + {\overline \Sigma}(\omega,{\bf k})}{{\cal D}(\omega,{\bf k})},\\
F^\dagger(\omega,{\bf k}) &=& -\frac{B_{\bf k} + \Pi^\dagger(\omega,{\bf k})}{{\cal D}(\omega,{\bf k})},\nonumber
\end{eqnarray}
where
\begin{eqnarray}
\label{d}
{\cal D}(\omega,{\bf k}) &=& (\omega+i\delta)^2 - \epsilon_{0\bf k}^2 - \Omega(\omega,{\bf k}),\\
\label{o}
\Omega(\omega,{\bf k}) &=& E_{\bf k}(\Sigma + \overline{\Sigma}) - B_{\bf k}\Pi - B^*_{\bf k}\Pi^\dagger - (\omega + i\delta)(\Sigma - \overline{\Sigma}) - \Pi\Pi^\dagger + \Sigma \overline{\Sigma}
\end{eqnarray}
and $\epsilon_{0\bf k}$ is given by Eq.~(\ref{spec1}). The quantity $\Omega(\omega,{\bf k})$ describing the spin-wave spectrum renormalization is considered below. The last two terms in Eq.~(\ref{o}) give corrections of at least second order in $1/S$.

\section{Renormalization of the energy and the real part of the spectrum}
\label{realp}

The classical ground state of the model \eqref{ham} is continuously degenerate: magnetization can be oriented in any direction. It is well known, however, that quantum fluctuations lead to anisotropic corrections to the energy selecting a limiting number of states. \cite{tes,kef_an} Intrinsic anisotropy of the dipolar interaction is the origin of this effect. Anisotropic part of the first $1/S$-correction to the energy $E_0$ has the form
\begin{eqnarray}
\label{an}
\frac{\Delta E}{\mathfrak N} &=& C \frac{S^2\omega_0^2}{4D}(\gamma_x^2\gamma_y^2 + \gamma_x^2\gamma_z^2 + \gamma_y^2\gamma_z^2 ),\\
\label{c}
C &=& \frac{D}{\omega_0^2 \mathfrak N} \sum_{\bf q}\frac{\left( Q_{\bf q}^{xx} - Q_{\bf q}^{yy} \right)^2 - 4 \left( Q_{\bf q}^{xy} \right)^2 }{4\epsilon_{\bf q}},
\end{eqnarray}
where $\gamma_i$ are direction cosines of the magnetization and components of the dipolar tensor in Eq.~(\ref{c}) are taken relative to cubic axes. The constant $C$ can be calculated numerically and one obtains in accordance with Refs.~\cite{tes,kef_an} that it is positive for simple ($C\approx0.012$) and negative for face-centered ($C\approx-0.005$) and body-centered ($C\approx-0.04$) cubic lattices. Then, an edge of the cube is easy direction in the simple cubic lattice whereas a body diagonal of the cube is easy direction in face- and body-centered cubic lattices.

It is well known that fluctuations leading to anisotropic corrections to the energy which low the energy symmetry to a discrete one naturally lead also to a gap in the spectrum. To mention only a few, examples are antiferromagnet containing two coupled antiferromagnetic sublattices \cite{shender} and square planar rotator model with dipolar interaction. \cite{anis_it} As we show in Refs.~\cite{idipfm,i2d}, it is also the case in the considered 3D FM and in 2D FM with dipolar forces.

\begin{figure}
\centering
\includegraphics[scale=0.3]{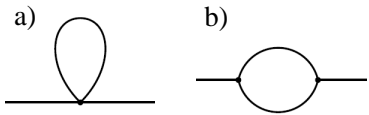}
\caption{Diagrams for self-energy parts of the first order in $1/S$.
\label{diagfig}} 
\end{figure}

Renormalization of $\epsilon_{0\bf k}$ in the first order in $1/S$ comes from two diagrams shown in Fig.~\ref{diagfig}. Both diagrams contribute to renormalization of the real part of the spectrum and only the loop diagram (b) leads to magnon damping. Renormalization of the real part of the spectrum was discussed in detail in our previous paper Ref.~\cite{idipfm}. It is demonstrated there that renormalized spectrum has the form
\begin{eqnarray}
\label{spec}
\epsilon_{\bf k} &=& 
\sqrt{
\epsilon_{0\bf k}^2
+
2\frac{D\Delta^2}{S\omega_0}k^2
+ 
\Delta^2\sin^2\theta_{\bf k}
},\\
\label{delta}
\Delta^2 &=& \left\{
\begin{aligned}
& Y\frac{S^2\omega_0^3}{D}, & T&\ll\sqrt{SD\omega_0},\\
& \frac{1}{2^8}\frac{S^2\omega_0^3T}{D\sqrt{SD\omega_0}}, & T&\gg\sqrt{SD\omega_0},
\end{aligned}
\right.
\end{eqnarray}
where $Y=C/2\approx0.006$ for simple cubic lattice and $Y=-3C/4$ for face- and body-centered cubic lattices. Notice that we present in Eq.~\eqref{spec} only corrections which change drastically the bare spectrum $\epsilon_{0\bf k}$ at $T\ll T_C$. The rest corrections observed in Ref.~\cite{idipfm} lead to only small renormalization of constants $D$ and $\omega_0$ in Eq.~\eqref{spec0}. In particular, the term in Eq.~\eqref{spec} proportional to $k^2$ makes the spectrum linear at sufficiently small $k$ and $\sin\theta_{\bf k}=0$ whereas $\epsilon_{0\bf k}\propto k^2$. The last term under the square root in Eq.~\eqref{spec} is the square of the gap value. Observation of the gap induced by dipolar forces resolves problems of infrared singularities of longitudinal spin susceptibility and corrections to the spin-wave stiffness in 3D FM \cite{idipfm}. Eq.~\eqref{spec} can be simplified as follows in three limiting cases which are considered below:
\begin{equation}
\label{specas}
\epsilon_{\bf k} \approx \left\{
\begin{aligned}
& Dk^2, & \sqrt{\frac{S\omega_0}{D}}\ll k&\ll 1,\\
& k\sqrt{SD\omega_0}\sin\theta_{\bf k}, & \frac{\Delta}{\sqrt{SD\omega_0}}\ll k&\ll \sqrt{\frac{S\omega_0}{D}},\\
& \Delta\sin\theta_{\bf k}, & k&\ll \frac{\Delta}{\sqrt{SD\omega_0}},
\end{aligned}
\right.
\end{equation}
where we assume in the second and the third lines that $\sin\theta_{\bf k}$ is not too small: $\sin\theta_{\bf k}\gg k\sqrt{D/S\omega_0}\ll1$.

\section{Magnon damping}
\label{md}

Imaginary part of the loop diagram shown in Fig.~\ref{diagfig}(b) can be obtained quite straightforwardly as it was done, for instance, in our previous paper \cite{i2d} devoted to 2D FM with dipolar interaction. Some details of these simple but tedious calculations together with general expression for ${\rm Im}\Omega(\omega,{\bf k})$ in the first order in $1/S$ can be found in Appendix. The value ${\rm Im}\Omega(\omega,{\bf k})$ is an odd function of $\omega$ and we calculate it for $\omega=\epsilon_{\bf k}$ only. According to Eq.~\eqref{d} the spin-wave damping $\Gamma_{\bf k}$ at momentum $\bf k$ is given in the first order in $1/S$ by the relation 
\begin{equation}
\label{dam}
\Gamma_{\bf k} = -\frac{{\rm Im}\Omega(\omega=\epsilon_{\bf k},{\bf k})}{2\epsilon_{\bf k}}.
\end{equation}
Strictly speaking, one has to put the bare spectrum $\epsilon_{0\bf k}$ in Eq.~\eqref{dam} instead of $\epsilon_{\bf k}$ in the first order in $1/S$. It is shown in Appendix that ${\rm Im}\Omega(\omega=\epsilon_{0\bf k},{\bf k})\equiv0$ at $\sin\theta_{\bf k}=0$. But, as we show below, the damping diverges as $k\to0$ at $T\ne0$ and $\sin\theta_{\bf k}\ne0$. This divergence signifies that the spectrum cannot be found in conventional way as a series of integer powers of $1/S$. However one may try to find a solution of the equation ${\cal D}(\omega,{\bf k})=0$ self-consistently using general expressions for self-energy parts obtained within the first few orders in $1/S$. It was the way in which the real part of the spectrum \eqref{spec} was obtained in Ref.~\cite{idipfm}. The only assumption we made there is that $\Gamma_{\bf k}\ll\epsilon_{\bf k}$ at $k\gg\Delta/\sqrt{SD\omega_0}$ (that is really the case as results show presented below). Notice, in particular, that the spectrum \eqref{spec} is of the order of $\sqrt S$ at $k=0$, $\sin\theta_{\bf k}\ne0$ and $T\ll\sqrt{SD\omega_0}$, while the bare spectrum is of the order of $S^1$ and corrections to it of the first order in $1/S$ are of the order of $S^0$. Such dependence of the spectrum on fractional powers of $S$ is the result of the self-consistent procedure being used.

We obtain below $\Gamma_{\bf k}$ self-consistently using the expression for self-energy parts in the first order in $1/S$ and assuming that $\epsilon_{\bf k}$ is the renormalized spectrum given by Eq.~\eqref{spec}. In particular, the gap in the spectrum screens the divergence of $\Gamma_{\bf k}$ obtained in the first order of the conventional $1/S$ expansion. The results presented below for $k\gg \sqrt{S\omega_0/D}$ are valid for all $\theta_{\bf k}$ whereas the damping was found at $k\ll \sqrt{S\omega_0/D}$ under assumption that $\sin\theta_{\bf k}\gg k\sqrt{D/S\omega_0}\ll1$. The analysis becomes more complicated at $k\ll \sqrt{S\omega_0/D}$ in the narrow interval of angles given by inequality $\sin\theta_{\bf k}\alt k\sqrt{D/S\omega_0}\ll1$. The damping is expected to be small at such $\theta_{\bf k}$ because $\Gamma_{\bf k}=0$ at $\sin\theta_{\bf k}=0$ in the first order in $1/S$, as it is noted above. It is shown below that expressions for self-energy parts of higher orders in $1/S$ involved in the self-consistent calculations give negligibly small contribution to the result by the parameter $S\omega_0/D\ll1$. It is convenient to consider two regimes, $T=0$ and $T\gg S\omega_0$, in which only quantum and thermal fluctuations determine the damping, respectively.
 
\subsection{$T=0$}
\label{mdst}

Decay processes \eqref{split} lead to magnon damping at $T=0$ which has the form
\begin{equation}
\label{dam0}
\Gamma_{\bf k} = 
\left\{
\begin{aligned}
& \frac{1}{2^9\pi}\frac{S\omega_0^2}{D}k(4-3\sin^2\theta_{\bf k})\sin^2\theta_{\bf k}, & k&\gg\sqrt{\frac{S\omega_0}{D}},\\
& \frac{1}{672\sqrt2\pi}\omega_0\left(\frac{\epsilon_{\bf k}}{D}\right)^{3/2}\sin^2\theta_{\bf k}, & k&\ll\sqrt{\frac{S\omega_0}{D}},
\end{aligned}
\right.
\end{equation}
where $\epsilon_{\bf k}$ is given by Eq.~\eqref{spec} and it is taken into account that $\epsilon_{\bf k}\approx Dk^2$ at $k\gg\sqrt{S\omega_0/D}$. It is seen from Eqs.~\eqref{specas} and \eqref{dam0} that $\Gamma_{\bf k}$ is much smaller than $\epsilon_{\bf k}$ by the parameter $S\omega_0/D\ll1$ and magnons are well-defined quasiparticles at $T=0$. It is seen from Eqs.~\eqref{specas} and \eqref{dam0} that for a given $\theta_{\bf k}$ the damping decreases monotonically as $k$ decreases in the interval $\Delta/\sqrt{SD\omega_0}\ll k\ll1$ and it is flat at $k\ll\Delta/\sqrt{SD\omega_0}$.

\subsection{$T\gg S\omega_0$}
\label{mdlt}

\begin{figure}
\centering
\includegraphics[scale=0.9]{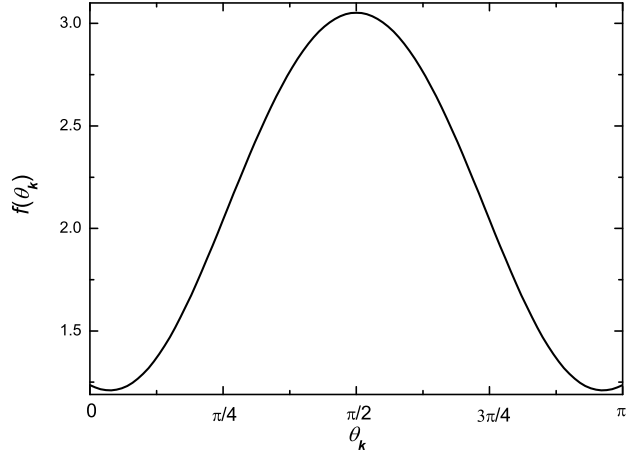}
\caption{Function $f(\theta_{\bf k})$ is shown that is given by Eq.~\eqref{f}.
\label{ffig}} 
\end{figure}

As we note above, the damping diverges at $T\ne0$ as $\Gamma_{{\bf k}\to0}\propto T\omega_0^{5/2}/\epsilon_{0\bf k}$ but one obtains a finite $\Gamma_{\bf k}$ as a result of self-consistent calculation.
We obtain assuming that $T\gg\epsilon_{\bf k}$ after tedious calculations some detail of which are presented in Appendix
\begin{widetext}
\begin{equation}
\label{damne0}
\Gamma_{\bf k} = 
\begin{cases}
\displaystyle \frac{1}{2^7\pi}\frac{TS\omega_0^2}{D^2k}
\left(
(4-3\sin^2\theta_{\bf k}) \ln\left(\frac{Dk^2}{S\omega_0}\right)
+
\frac{16\cos^2\theta_{\bf k}}{1+|\cos\theta_{\bf k}|} 
\ln\left(\frac{S\omega_0}{\Delta} \right)
\right)\sin^2\theta_{\bf k}, &   
\displaystyle \sqrt{\frac{S\omega_0}{D}}\ll k\ll1,\\
\displaystyle \frac{1}{2^7\pi^2}\frac{T\omega_0}{\epsilon_{\bf k}}
\left(\frac{S\omega_0}{D}\right)^{3/2}\ln\left(\frac{\epsilon_{\bf k}}{\Delta\sin\theta_{\bf k}}\right) f(\theta_{\bf k})
\sin^{5/2}\theta_{\bf k}, & 
\displaystyle \frac{\Delta}{\sqrt{SD\omega_0}}\ll k\ll\sqrt{\frac{S\omega_0}{D}},\\
\displaystyle \frac{1}{192\pi}\frac{TS^2\omega_0^3}{D}\frac{k}{\epsilon_{\bf k}^2}
(2-\sin^2\theta_{\bf k})\sin^2\theta_{\bf k}, & 
\displaystyle {\cal K}
\ll k \ll\frac{\Delta}{\sqrt{SD\omega_0}},\\
\displaystyle \frac{1}{168\sqrt2\pi} \frac{T\omega_0}{D}\sqrt{\frac{\epsilon_{\bf k}}{D}}\sin^2\theta_{\bf k}, & \displaystyle k\ll {\cal K},
\end{cases}
\end{equation}
\end{widetext}
where ${\cal K} = \max\{\Delta^{5/2}/(S^2\omega_0^2\sqrt D),\Delta/\sqrt{TD}\}$,
\begin{equation}
\label{f}
f(\theta_{\bf k}) = 
\int_0^{2\pi}d\theta
\frac{\sin^2(\theta-\theta_{\bf k})}{\sqrt{\sin\theta_{\bf k}-2\sin\theta}}
H(\sin\theta_{\bf k}-2\sin\theta)
\end{equation}
and $H(x)$ is the Heaviside step function. As is seen from its graphic shown in Fig.~\ref{ffig}, values of the function $f(\theta_{\bf k})$ lie between 1.2 and 3.1. Both decay \eqref{split} and confluence \eqref{conf} processes contribute to the damping at $k\gg \sqrt{S\omega_0/D}$ whereas confluence and decay processes dominate at ${\cal K}\ll k\ll\sqrt{S\omega_0/D}$ and $k\ll{\cal K}$, respectively. 

\begin{figure}
\centering
\includegraphics[scale=0.5]{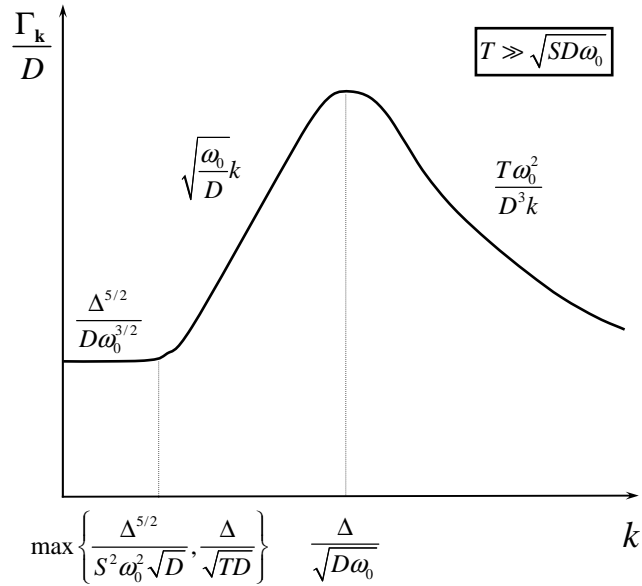}
\caption{Sketch of the damping $\Gamma_{\bf k}$ at $T\gg\sqrt{SD\omega_0}$ given by Eq.~\eqref{damne0}. Thermal fluctuations are responsible for the peak at $k\sim\Delta/\sqrt{SD\omega_0}$: damping at $T=0$ given by Eq.~\eqref{dam0} rises monotonically as $k$ increases.
\label{dam3d}} 
\end{figure}

It is seen from Eqs.~\eqref{delta}, \eqref{specas} and \eqref{damne0} that the damping rises as $T\omega_0^2/(D^2k)$ upon the momentum decreasing up to $k\sim \Delta/\sqrt{SD\omega_0}$, it decreases linearly as $\sqrt{D\omega_0}k$ at ${\cal K}\ll k\ll \Delta/\sqrt{SD\omega_0}$ and it is flat at $k\ll {\cal K}$. Thus there is a peak in the damping at $k\sim \Delta/\sqrt{SD\omega_0}$. We draw $\Gamma_{\bf k}$ schematically in Fig.~\ref{dam3d} using Eq.~\eqref{damne0} in the particular case of $T\gg\sqrt{SD\omega_0}$ when $\Delta$ is given by the second line in Eq.~\eqref{delta}. 

\begin{figure}
\centering
\includegraphics[scale=0.8]{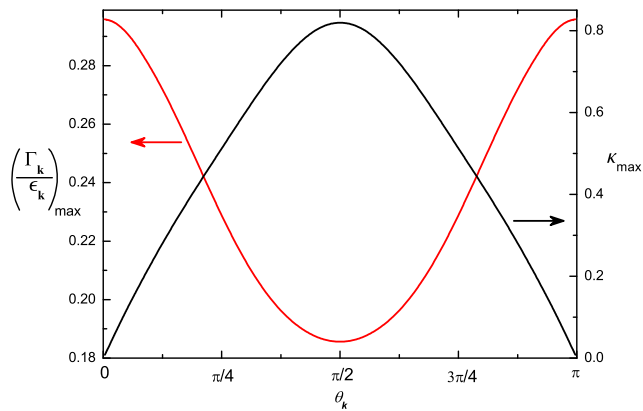}
\caption{(Color online). The peak height $(\Gamma_{\bf k}/\epsilon_{\bf k})_{\max}$ and position $k_{\max}=\kappa_{\max}\Delta/\sqrt{SD\omega_0}$ of the ratio $\Gamma_{\bf k}/\epsilon_{\bf k}$ at $T\gg\sqrt{SD\omega_0}$ as functions of $\theta_{\bf k}$.
\label{max}} 
\end{figure}

The ratio $\Gamma_{\bf k}/\epsilon_{\bf k}$ follows qualitatively the behavior of $\Gamma_{\bf k}$ and has a maximum at $k\sim \Delta/\sqrt{SD\omega_0}$ too. The peak height can be estimated from the third line in Eq.~\eqref{damne0} at $k\sim \Delta/\sqrt{SD\omega_0}$ that gives $\Gamma_{\bf k}/\epsilon_{\bf k}\sim T/\sqrt{SD\omega_0}\ll1$ at $T\ll\sqrt{SD\omega_0}$ and $\Gamma_{\bf k}/\epsilon_{\bf k}\sim1$ at $T\gg\sqrt{SD\omega_0}$. The peak height cannot be calculated analytically. We perform numerical integration to find $\Gamma_{\bf k}/\epsilon_{\bf k}$ at $k\sim \Delta/\sqrt{SD\omega_0}$ in the particularly interesting case of $T\gg\sqrt{SD\omega_0}$ which results are shown in Fig.~\ref{peaks} (see Appendix for detail). Asymptotics obtained from Eqs.~\eqref{spec} and \eqref{damne0} are also shown in Fig.~\ref{peaks} by dashed lines. As the asymptotic for $\Delta/\sqrt{SD\omega_0}\ll k\ll\sqrt{S\omega_0/D}$ was found with the "logarithmic" precision, we have introduced a factor of the order of unity under the logarithm in order to improve the quantitative agreement in the near vicinity of the peak between results of the analytical consideration and numerical integration. In particular, the factor introduced was equal to 1.03, 1.45 and 2.25 for $\theta_{\bf k}=\pi/2$, $\pi/4$ and $\pi/8$, respectively. It is seen from Fig.~\ref{peaks} that the peak becomes sharper and higher upon decreasing of $\sin\theta_{\bf k}$ and its position moves to smaller momenta. Characteristics of the peak, its height and position, as functions of $\theta_{\bf k}$ are presented in Fig.~\ref{max}. It is seen that the value of the peak height increases from 0.185 at $\theta_{\bf k}=\pi/2$ to 0.296 at $\sin\theta_{\bf k}\ll 1$. The rapid decrease to zero at $\sin\theta_{\bf k}\alt k\sqrt{D/S\omega_0}\ll1$ discussed above is not shown in Fig.~\ref{max}.

\begin{figure}
\centering
\includegraphics[scale=0.9]{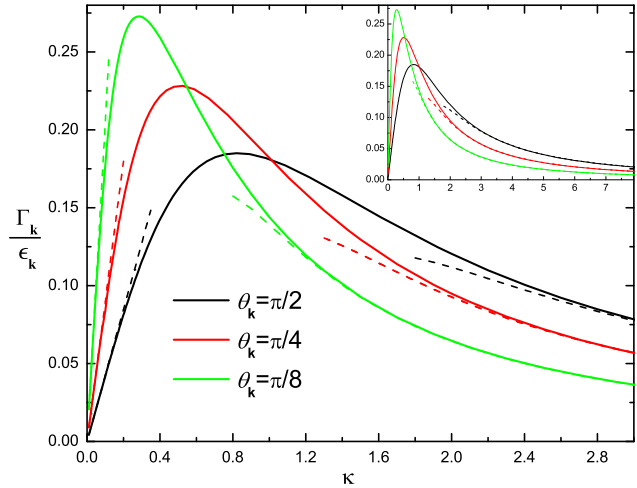}
\caption{(Color online). The ratio is shown of the magnon damping $\Gamma_{\bf k}$ and the real part of the spectrum $\epsilon_{\bf k}$ at $T\gg\sqrt{SD\omega_0}$ as a function of reduced momentum $\kappa=k\sqrt{SD\omega_0}/\Delta$ in the vicinity of the peak for three particular values of $\theta_{\bf k}$. Asymptotics at $k\ll\Delta/\sqrt{SD\omega_0}$ and $k\gg\Delta/\sqrt{SD\omega_0}$ are shown by dashed lines which are obtained from Eqs.~\eqref{specas} and \eqref{damne0}.
\label{peaks}} 
\end{figure}

\section{Discussion}
\label{disc}

{\it Classical-spin limit.}---It should be noted that anomalously large damping obtained above can be seen both in quantum and classical FMs because thermal fluctuations are responsible for it. Really, we conclude from Eqs.~\eqref{spec} and \eqref{damne0} taking into account that $T_C\propto S^2J$ at large $S$ that at fixed ratio $T/T_C$ one has $\epsilon_{\bf k},\Gamma_{\bf k}\sim S$. Thus both $\Gamma_{\bf k}/\epsilon_{\bf k}$ and the peak position are $S$-independent. As it is explained in Refs.~\cite{har,loly,i2d}, the spectrum and corrections to it in the classical limit can be obtained from expressions found above by multiplying them by $S$ and taking the limit
\begin{equation}
\label{lim}
S\to\infty, \quad \hbar\to0, \quad J,\omega_0\to0
\end{equation}
assuming that 
\begin{equation}
\label{ass}
\hbar S = {\rm const}, \quad JS^2 = {\rm const}, \quad \omega_0S^2 = {\rm const}.
\end{equation}
Quantum corrections (i.e., $T$-independent ones) die out as a result of this procedure.

{\it Role of the long-range character of the dipolar forces.}---It should be stressed that the long-range nature of dipolar forces is the origin of such a large damping obtained. It can easily be shown by restricting dipolar forces radius of action to a few lattice spacings. Such a restriction leads to small damping because the value of three-magnon vertex \eqref{h3} becomes of the order of $\omega_0k^2$ if $k_{1,2,3}\sim k\ll1$ rather than $\omega_0$. It is well known that small short-range interactions of other types lead to small damping as well. \cite{chub}

{\it Comparison with 2D FM with dipolar forces.}---Interestingly, quantum and thermal fluctuations lead to smaller $\Gamma_{\bf k}$ (compared to $\epsilon_{\bf k}$) in lower dimension 2D FM on square lattice with dipolar forces. It is demonstrated in Ref.~\cite{i2d} that thermal fluctuations lead also to a peak in $\Gamma_{\bf k}/\epsilon_{\bf k}$ at small $k$ in 2D FM which height, however, is of the order of $T/T_C\ll1$ for $S\sim1$ and reaches the value of 0.16 for $S\gg1$. The origin of the greater role which play fluctuations in higher dimension is that $Q_{\bf k}^{\alpha\beta}\propto\omega_0k$ in 2D FM whereas $Q_{\bf k}^{\alpha\beta}\propto\omega_0$ in 3D FM that leads to larger three-particle vertex in 3D FM.

\begin{figure}
\centering
\includegraphics[scale=0.6]{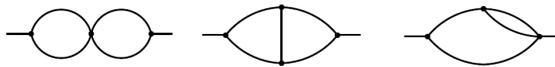}
\caption{Some diagrams of higher order in $1/S$.
\label{so}} 
\end{figure}

{\it Higher-order diagrams.}---The interesting property of the results discussed here is that higher-order diagrams some of which are presented in Fig.~\ref{so}, being included in the self-consistent calculations, give a negligibly small contribution by the small parameter $S\omega_0/D$. The origin of this fact is that singularities of higher-order diagrams are screened by the gap and all vertexes are small being by the order of magnitude not larger than $\omega_0$ when corresponding momenta are much smaller than $\sqrt{S\omega_0/D}$.

{\it Experimental verification.}---The main obstacle for experimental observation of anomalous damping in FMs discussed above is magnetocrystalline anisotropy which contribution to the gap in the spectrum $\Delta_a$ is much larger than $\Delta$ at $T\ll T_C$ in the majority of materials. The isotropic gap $\Delta_a$, in contrast to anisotropic one of the dipolar origin $\Delta\sin\theta_{\bf k}$, leads to exponential decay of the damping 
\begin{equation}
\label{expdam}
\Gamma_{\bf k}\propto\exp\left(-\frac{\Delta_a^2}{4TDk^2}\right) \quad\mbox{  at  }\quad k<\frac{\Delta_a}{\sqrt{TD}}
\end{equation}
screening the effect we discuss. However, there is a class of insulating FM materials which magnetism is due to magnetic ions in S-state and which, as a consequence, have very small magnetocrystalline anisotropy. The most suitable compounds for corresponding experiments seem EuS ($T_C=16.6$ K, $S=7/2$, $D\approx1.68$ K, $\omega_0\approx0.58$ K, $\Delta_a\approx0.005$ K) \cite{eus1,*eus2} and $\rm CdCr_2Se_4$ ($T_C=130$ K, $S=3/2$, $D\approx3.9$ K, $\omega_0\approx0.3$ K, $\Delta_a\approx0.002$ K) \cite{cd1,*cd2}. It should be noted that because $S\omega_0>D$, asymptotics \eqref{dam0} and \eqref{damne0} for $\Gamma_{\bf k}$ are not valid for EuS. However, calculations of $\Gamma_{\bf k}$ at $k\sim\Delta/\sqrt{SD\omega_0}$ and $\sin\theta_{\bf k}\sim1$ showing the peak remain valid for EuS. Particular estimations demonstrate that the maximum of the damping in EuS and $\rm CdCr_2Se_4$ at $\theta_{\bf k}=\pi/2$ and $T=0.1\div0.2T_C$ is at $k\approx0.01 \mbox{\AA}^{-1}$ and $0.002 \mbox{\AA}^{-1}$, respectively.

\section{Conclusion}
\label{con}

To conclude, we demonstrate by the example of 3D Heisenberg FM with arbitrary small dipolar forces at $T\ll T_C$ that long-range interaction in a system {\it can} lead to a very strong interaction between long-wavelength quasiparticles and make them heavily damped. We show that magnons are well-defined quasiparticles in 3D FM at $T=0$ which damping is given by Eq.~\eqref{dam0}. Thermal fluctuations give the main contribution to the damping at $T\gg S\omega_0$ which is given by Eq.~\eqref{damne0} and which is also sketched in Fig.~\ref{dam3d}. There is a peak in the damping at such $T$ at $k_{max}\sim \Delta/\sqrt{SD\omega_0}$. There is also a peak at $k\sim k_{max}$ in the ratio $\Gamma_{\bf k}/\epsilon_{\bf k}$ which height becomes of the order of unity when $T\gg\sqrt{SD\omega_0}$. The ratio $\Gamma_{\bf k}/\epsilon_{\bf k}$ is shown in Fig.~\ref{peaks} in the vicinity of the peak. The peak height and position are plotted in Fig.~\ref{max}. The maximum value of $\Gamma_{\bf k}/\epsilon_{\bf k}$ approaches 0.3 at small $\sin\theta_{\bf k}$. It should be noted that only small fraction of magnons with $\epsilon_{\bf k}<T$ appears to be overdamped. That is why one could expect small influence from them on the system properties. However, the observed suppression of long-wavelength magnons is remarkable because it contradicts expectation of the quasiparticle concept.

Because thermal fluctuations are responsible for the peaks in $\Gamma_{\bf k}$ and $\Gamma_{\bf k}/\epsilon_{\bf k}$, they can be seen both in quantum and classical FMs. We demonstrate that it is the long-range nature of dipolar forces that is responsible for the anomalously large damping observed in 3D FM: magnons are well-defined quasiparticles if one restricts dipolar forces radius of action to a few lattice spacings. We make a counter intuitive conclusion that dipolar forces lead to smaller magnon interaction in lower dimension 2D FM discussed in our previous paper \cite{idipfm}, where the peak in the ratio $\Gamma_{\bf k}/\epsilon_{\bf k}$ is also observed which height, however, is of the order of $T/T_C\ll1$ for $S\sim1$ and it reaches 0.16 for $S\gg1$. We show that magnetocrystalline anisotropy leading to isotopic contribution to the gap $\Delta_a$ screens the enhancement of the damping we discuss because, in contrast to anisotropic gap $\Delta\sin\theta_{\bf k}$ of the dipolar origin, it leads to exponential decay of the damping at small $k$ (see Eq.~\eqref{expdam}). Seemingly surprising fact that such a remarkable anomaly of the damping was not obtained before can be explained by small value of $k_{max}$ and by large value of $\Delta_a$ as compared to $\Delta$ in the majority of materials. We show that this effect can be observed experimentally in materials with small magnetocrystalline anisotropy the most suitable of which seem EuS and $\rm CdCr_2Se_4$.

\begin{acknowledgments}

This work was supported by President of Russian Federation (grant MK-329.2010.2), RFBR grant 09-02-00229, and Programs "Quantum Macrophysics", "Strongly correlated electrons in semiconductors, metals, superconductors and magnetic materials" and "Neutron Research of Solids".

\end{acknowledgments}

\appendix

\section{Details of the damping calculation}

Taking into account that summation over small momenta is essential in calculation of the damping at $k\ll1$ and $T\ll T_C$ one has for the contribution to $\Omega(i\omega,{\bf k})$ from the loop diagram shown in Fig.~\ref{diagfig}(b)
\begin{subequations}
\label{o3}
\begin{eqnarray}
\Omega(i\omega,{\bf k}) &=& -Dk^2\frac{S\omega_0^2}{4\mathfrak N} 
T\sum_{\omega_1+\omega_2=\omega}\sum_{{\bf q}_1+{\bf q}_2={\bf k}} 
\frac{1}{[(i\omega_1)^2 - \epsilon_1^2][(i\omega_2)^2 - \epsilon_2^2]} \nonumber\\
&&\times\{
\sin^22\theta_{\bf k}(B_1B_2^* + E_1E_2 + \omega_1\omega_2)\\
&&{}-2\sin2\theta_{\bf k}( S\omega_0Dq_1^2\sin2\theta_1\sin^2\theta_2\sin(\phi_1-\phi_2)\sin(\phi_2-\phi_{\bf k}) \nonumber\\ 
&&{}- (Dq_1^2Dq_2^2+\omega_1\omega_2)\sin2\theta_1\cos(\phi_1-\phi_{\bf k}))\\
&&{} + ( 2Dq_1^2Dq_2^2\sin^22\theta_1 + S\omega_0Dq_1^2\sin^2\theta_2\sin^22\theta_1\nonumber\\  
&&{}+ (Dq_1^2Dq_2^2 - \omega_1\omega_2)\sin2\theta_1\sin2\theta_2\cos(\phi_1-\phi_2) )\}\\
&&{}-\frac{S^2\omega_0^3}{4\mathfrak N}\sin^2\theta_{\bf k} 
T\sum_{\omega_1+\omega_2=\omega}\sum_{{\bf q}_1+{\bf q}_2={\bf k}} 
\frac{1}{[(i\omega_1)^2 - \epsilon_1^2][(i\omega_2)^2 - \epsilon_2^2]} \nonumber\\
&&\times\{
Dq_1^2Dq_2^2\sin^22\theta_1 + S\omega_0Dq_1^2\sin^2\theta_2\sin^22\theta_1\cos^2(\phi_2-\phi_{\bf k})\nonumber\\
&&{} + Dq_1^2Dq_2^2\sin2\theta_1\sin2\theta_2\sin(\phi_1-\phi_{\bf k})\sin(\phi_2-\phi_{\bf k})\nonumber\\
&&{} - \omega_1\omega_2\sin2\theta_1\sin2\theta_2\cos(\phi_1-\phi_{\bf k})\cos(\phi_2-\phi_{\bf k}) \}\\
&&{} -i\omega\frac{S\omega_0^2}{2\mathfrak N} 
T\sum_{\omega_1+\omega_2=\omega}\sum_{{\bf q}_1+{\bf q}_2={\bf k}} 
\frac{i\omega_1}{[(i\omega_1)^2 - \epsilon_1^2][(i\omega_2)^2 - \epsilon_2^2]} \nonumber\\
&&{}\times \{
Dq_2^2\sin2\theta_2 (\sin2\theta_2 + \sin2\theta_1\cos(\phi_1-\phi_2))\\
&&{} + \sin2\theta_{\bf k} (Dq_2^2\sin2\theta_2\cos(\phi_2-\phi_{\bf k}) - Dq_2^2\sin2\theta_1\cos(\phi_1-\phi_{\bf k})\nonumber\\ 
&&{} - S\omega_0\sin2\theta_1\sin^2\theta_2\cos(\phi_2-\phi_{\bf k})\cos(\phi_2-\phi_1)) \}.
\end{eqnarray}
\end{subequations}
We obtain after summation over imaginary frequencies and analytical continuation on $\omega$ from an imaginary axis to the real one
\begin{subequations}
\label{imsum}
\begin{eqnarray}
&&{\rm Im}\left(T\sum_{\omega_1} \frac{1}{[(i\omega_1)^2 - \epsilon_1^2][(i\omega_1 - i\omega)^2 - \epsilon_2^2]}\right)
= \nonumber\\
&& \frac{\pi}{4\epsilon_1\epsilon_2}(1+N_1+N_2){\rm sgn}(\omega)\delta(|\omega|-\epsilon_1-\epsilon_2)
- \frac{\pi}{4\epsilon_1\epsilon_2}(N_1-N_2) ( \delta(\omega-\epsilon_1+\epsilon_2) -  \delta(\omega+\epsilon_1-\epsilon_2) ),
\\
&& {\rm Im}\left(T\sum_{\omega_1} \frac{(i\omega_1)(i\omega-i\omega_1)}{[(i\omega_1)^2 - \epsilon_1^2][(i\omega_1 - i\omega)^2 - \epsilon_2^2]} \right)
= \nonumber\\
&& \frac{\pi}{4}(1+N_1+N_2){\rm sgn}(\omega)\delta(|\omega|-\epsilon_1-\epsilon_2)
+ \frac{\pi}{4}(N_1-N_2) ( \delta(\omega-\epsilon_1+\epsilon_2) -  \delta(\omega+\epsilon_1-\epsilon_2) ),
\\
&& {\rm Im}\left(T\sum_{\omega_1} \frac{i\omega_1}{[(i\omega_1)^2 - \epsilon_1^2][(i\omega_1 - i\omega)^2 - \epsilon_2^2]}\right) =
\nonumber\\ 
&& \frac{\pi}{4\epsilon_2}(1+N_1+N_2)\delta(|\omega|-\epsilon_1-\epsilon_2)
- \frac{\pi}{4\epsilon_2}(N_1-N_2) ( \delta(\omega-\epsilon_1+\epsilon_2) + \delta(\omega+\epsilon_1-\epsilon_2) ).
\end{eqnarray}
\end{subequations}
The first and the second terms in Eqs.~\eqref{imsum} describe magnon decay and confluence processes, respectively. It can be straightforwardly shown using Eqs.~\eqref{o3} and \eqref{imsum} that ${\rm Im}\Omega(\omega,{\bf k})=0$ at $\sin\theta_{\bf k}=0$ if $\epsilon_{1,2}$ and $\epsilon_{\bf k}$ are bare spectra given by Eq.~\eqref{spec0}.

\subsection{$T=0$}

Only decay processes contribute to the damping at $T=0$ because $N_1=N_2=0$ in Eqs.~\eqref{imsum}. The main contribution at $1\gg k\gg\sqrt{S\omega_0/D}$ comes from terms (c) and (e) in Eq.~\eqref{o3}. Summation over momentum $q_1\sim k$ is essential in this case that leads to the first line in Eq.~\eqref{dam0}.

The main contribution at $k\ll\sqrt{S\omega_0/D}$ comes only from term (d) in Eq.~\eqref{o3}. Summation over $q_1\gg k$ and $\sin\theta_1\ll q_1\sqrt{D/(S\omega_0)}$ is essential that leads to the second line in Eq.~\eqref{dam0}, where $\epsilon_{\bf k}$ should be replaced by $\epsilon_{0\bf k}$ in the first order in $1/S$. 

Although there is no problem with infrared singularities at $T=0$, one has to go beyond the first order in $1/S$ to obtain correct expressions for the damping. It is the consequence of the fact that the bare spectrum is renormalized greatly by the first $1/S$ corrections at small momenta. The easiest way to do it in the present case is to find the damping self-consistently by using "dressed" Green's functions in the loop diagram and showing that higher order diagrams are negligible some of which are shown in Fig.~\ref{so}. Self-energy parts in numerators of Green's function \eqref{gf} are negligible at $k\ll1$ and $T\ll T_C$ and one can use Eq.~\eqref{o3} for the self-consistent calculation assuming that $\epsilon_{1,2}$ and $\epsilon_{\bf k}$ are renormalized spectra given by Eq.~\eqref{spec}. As a result one leads to Eq.~\eqref{dam0} for the damping.

\subsection{$T\gg S\omega_0$}

To simplify the consideration we assume that $T\gg\epsilon_{\bf k}$ that allows to replace $N_{1,2}$ in Eqs.~\eqref{imsum} by $T/\epsilon_{1,2}$ to obtain the main contribution to the damping. Terms (c) and (e) in Eq.~\eqref{o3} play the main role at $k\gg\sqrt{S\omega_0/D}$ and one leads to the first line in Eq.~\eqref{damne0} after self-consistent calculations. Notice that ${\rm Im}\Omega(\omega,{\bf k})$ has a logarithmic singularity in the first order in $1/S$ (the factor $\ln(S\omega_0/\Delta)$ diverges in Eq.~\eqref{damne0} as $\Delta\to0$) that is screened by the gap in the self-consistent procedure.

The main contribution at $k\ll\sqrt{S\omega_0/D}$ comes from term (d) in Eq.~\eqref{o3}. It can be readily shown that the energy conservation law of the confluence process \eqref{conf} cannot be fulfilled in the limit of small $k$ due to the gap in the spectrum. It is easy to realize that the corresponding contribution becomes exponentially small at $k<\Delta/\sqrt{TD}$. In contrast, equality describing magnon decay \eqref{split} is fulfilled by momenta which are nearly parallel to magnetization: $q\gg k$ and $\sin\theta_{\bf q}\ll q\sqrt{D/(S\omega_0)}$. Confluence processes make the main contribution to imaginary part of term (d) in Eq.~\eqref{o3} at $\Delta/\sqrt{SD\omega_0}\ll k\ll\sqrt{S\omega_0/D}$. But their contribution decreases at $k<\Delta/\sqrt{SD\omega_0}$ as $k$ decreases and it becomes of the same order as that from the decay processes at $k\sim{\cal K} = \max\{\Delta^{5/2}/(S^2\omega_0^2\sqrt D),\Delta/\sqrt{TD}\}$. Term (d) in Eq.~\eqref{o3} can be represented in the form at ${\cal K}\ll k\ll\sqrt{S\omega_0/D}$
\begin{eqnarray}
\label{peak}
\Gamma_{\bf k} &=& 
\frac{4}{\pi^2} \frac{\Delta}{\kappa}\sin\theta_{\bf k}\int_0^\pi d\theta\int_0^\infty dq 
\frac{\sin^22\theta(2q^2+A^2(\sin^2\theta-2q^2))}{(2q^2+1)(q^2+\sin^2\theta)^{3/2}\sqrt{1-A^2}}
H(1-A^2),\\
A &=& \frac{\sqrt{1+\kappa^2}}{\kappa}\frac{\sqrt{q^2+\sin^2\theta}}{(2q^2+1)\sin\theta} + \frac{2q^2\cos\theta}{(2q^2+1)\sin\theta}\cot\theta_{\bf k},
\end{eqnarray}
where $\kappa=k\sqrt{SD\omega_0}/\Delta$ and $H(x)$ is the Heaviside step function. Integrals in Eq.~\eqref{peak} can be taken if $k\gg \Delta/\sqrt{SD\omega_0}$ and $k\ll \Delta/\sqrt{SD\omega_0}$ that leads to the second and the third lines in Eq.~\eqref{damne0}, respectively. The peak height which is located at $k\sim \Delta/\sqrt{SD\omega_0}$ cannot be calculated using Eq.~\eqref{peak} analytically. Results of numerical integration at $T\gg\sqrt{SD\omega_0}$ using Eq.~\eqref{peak} are shown in Fig.~\ref{peaks}.

Decay processes come into play at $k\ll{\cal K}$ in term (d) of Eq.~\eqref{o3}, where summation over $q_1\gg k$ and $\sin\theta_1\ll q_1\sqrt{D/(S\omega_0)}$ is essential. One leads to the last line in Eq.~\eqref{damne0} that is simply the last line in Eq.~\eqref{dam0} multiplied by $4T/\epsilon_{\bf k}$.

\bibliography{dam3dfm}

\end{document}